\documentclass[pre,twocolumn,showpacs]{revtex4}
\usepackage{epsfig}
\usepackage[nooneline]{subfigure}
\usepackage[]{natbib}
\usepackage{amsmath}
\usepackage{caption}
\usepackage{graphics}
\usepackage{color}
\usepackage{array}
\usepackage{float}

\definecolor{HUblau}{rgb}{0,0.275,0.525}
\definecolor{hellblau}{rgb}{0.92,0.96,1}
\definecolor{BIUor}{rgb}{0.7,0.35,0.18}
\definecolor{apric}{rgb}{1,0.88,0.75}
\definecolor{Black}{rgb}{0,0,0}
\definecolor{White}{rgb}{1,1,1}
\definecolor{rosa}{rgb}{1,.95,.93}

\begin{document}

\title{A no-go theorem for ergodicity and an Einstein relation}

\author{D. Froemberg}
\author{E. Barkai}
\affiliation{Department of Physics, Institute of Nanotechnology and Advanced Materials, Bar Ilan University, Ramat-Gan
52900, Israel}


\begin{abstract}
We provide a simple no-go theorem for ergodicity and the generalized Einstein relation for anomalous diffusion processes.
The theorem states that either ergodicity in the sense of equal time and ensemble averaged 
mean squared displacements (MSD) is broken, and/or the generalized Einstein relation for time averaged diffusivity 
and mobility is invalid, which is in complete contrast to normal diffusion processes.
We also give a general relation for the time averages of drift and MSD for \textit{ergodic} (in the MSD sense) 
anomalous diffusion processes, showing that the ratio of these quantities depends on the measurement time.
The L\'evy walk model is used to exemplify the no-go theorem.
\end{abstract}

\maketitle

\paragraph*{Introduction.}
Many processes in disordered systems exhibit anomalous diffusion, that is a
nonlinear time dependence of the ensemble average mean squared displacement (MSD) 
$\langle x^2\rangle \propto t^\nu$, with $0<\nu<1$ for subdiffusion and $1<\nu<2$
for superdiffusion (or enhanced diffusion) \cite{LWKlaf87, LWKlaf90, Bou90, MetzKlaf00}. 
%
For normal Brownian diffusion, the Einstein relation connects the fluctuations of 
an ensemble of particles with their mobility $\mu$ (which is the inverse friction) 
under an applied small constant force $F$,
\begin{equation}
D = \mu k_B T \label{ein0}
\end{equation}
with $D$ being the diffusion constant, $k_B$ the Boltzmann constant and $T$ the temperature 
\cite{Einstein05, Marc08}.
Thereby the mobility is defined through $\langle x_F\rangle = \mu F t$ in one dimension.
The Einstein relation (\ref{ein0}) implies
\begin{eqnarray}
\langle x_F(t)\rangle 
&=& \frac{\langle x_0^2(t)\rangle}{2k_B T} F,
\label{einst}
\end{eqnarray}
where $\langle\cdots\rangle$ denotes ensemble averaging and $x_F$ and $x_0$ 
are the particle position with or without applied force $F$, respectively.
Eq. (\ref{einst}) holds for normal and anomalous processes 
close to equilibrium in the limit $F\to 0$ and can be derived from linear response theory 
\cite{Bou90, BarFleu98, BeOsh02, Zvi09}. Due to its validity beyond normal processes
Eq. (\ref{einst}) is referred to as the generalized Einstein relation (GER).
Measurements of this fundamental relation for $\nu\neq 1$ were performed 
e.g. for subdiffusive systems by \cite{Gu96, Amblard96}.

In what follows we will use the notion of ergodicity in the MSD sense:
A process is called ergodic, if time averaged and ensemble averaged MSDs are equal,
\begin{equation}
\lim_{t\to \infty} \overline{\delta_0^2} (t,\Delta) = \langle x_0^2(\Delta)\rangle \label{erg}
\end{equation}
where $\Delta$ is a time lag (see further discussion below).
In the case of Brownian diffusion Eq. (\ref{erg}) holds due to the stationary increments of the process.
For certain models of anomalous diffusion ergodicity is violated \cite{Lub08, HeBur08}.

In this brief report we formulate a no-go theorem for ergodicity and GER 
for time averages that applies to all kinds of anomalous diffusions. 
We then formulate a new type of relation between time averaged drift and MSD for L\'evy walks 
where the ratio of these two quantities is time dependent.
We then show that a similar time dependence holds also rather generally 
for ergodic anomalous diffusion processes. 
As discussed at the end of this paper, the method of averaging, i.e. time versus ensemble, 
determines the ratio between fluctuation and drift and hence effective temperatures.

\paragraph*{A no-go-theorem for ergodicity breaking and the generalized Einstein relation for time averages.} 

We define the GER for time averages by substituting 
ensemble averages $\langle\cdots\rangle$ by time averages $\overline{\cdots}$ in Eq. (\ref{einst}),
that is
\begin{equation} 
\overline{\delta}_F = \frac{F \overline{\delta_0^2} }{2k_B T}.
\end{equation}
In the following we will adhere to the ensemble average of $\overline{\delta_0^2}$, 
i.e. the mean taken over many realizations of trajectories so that the respective 
GER becomes
\begin{equation} 
\langle \overline{\delta}_F\rangle = \frac{F\langle \overline{\delta_0^2} \rangle}{2k_B T}.
\label{TAeinst}
\end{equation}
The time averaged MSD along the trajectory $x_0(t^\prime)$ is defined as
\begin{eqnarray}
\overline{\delta_0^2}(t,\Delta) &=& 
\frac{1}{t-\Delta} \int_0^{t-\Delta} \left[ x_0(t^\prime + \Delta) - x_0(t^\prime) \right]^2\; dt^\prime,
\label{tamsd0}
\end{eqnarray}
where the measurement time $t$ is much larger than the lag time $\Delta$.
Analogously, the time averaged drift is defined as 
\begin{equation}
\overline{\delta_F}(t,\Delta) = \frac{1}{t-\Delta}\int_0^{t-\Delta}  
\left[ x_F(t^\prime+\Delta) - x_F(t^\prime) \right]  dt^\prime  .
\label{TAdrift}
\end{equation}
With these definitions at hand we assert the general statement that in any system 
that exhibits anomalous diffusion (sub- or enhanced diffusion) 
and where a (generalized) Einstein relation of the type Eq. (\ref{einst}) for the ensemble averages holds,
\textit{at least} one of the two properties is violated: 
either ergodicity in the MSD sense or the GER for time-averages Eq. (\ref{TAeinst}).
Thus, let $\langle x_0^2(t)\rangle = 2 D_{\nu} t^{\nu}$, $0<\nu$ and $\nu\neq 1$, 
and according to the GER for ensemble averages Eq. (\ref{einst})
$\langle x_F(t)\rangle = F D_{\nu} t^{\nu}/(k_B T)$, where $D_\nu$ is a generalized diffusion constant.
Then, with Eq. (\ref{TAdrift}) we find
\begin{eqnarray}
\langle \overline{\delta}_F(t,\Delta)\rangle &=& \frac{F D_{\nu}}{k_B T} \frac{1}{(t-\Delta)} 
\int_0^t \left[ (t^\prime+\Delta)^{\nu} - t^{\prime \nu}\right]dt^\prime \nonumber \\
&=& \frac{F D_{\nu}\left[ (t+\Delta)^{1+\nu} - \Delta^{1+\nu} - t^{1+\nu}\right]}{k_B T(t-\Delta)(1+\nu)}
\end{eqnarray}
and for $\Delta \ll t$ in leading order
\begin{eqnarray}
\langle \overline{\delta}_F(t,\Delta)\rangle
&\sim& \frac{F D_{\nu}}{k_B T} \Delta t^{\nu-1}. \label{EAdrift0}
\end{eqnarray}
The GER for the time averages Eq. (\ref{TAeinst}) requires the time averaged MSD
\begin{eqnarray}
\langle \overline{\delta_0^2}(t,\Delta)\rangle
&=& \frac{2k_B T\langle \overline{\delta}_F(t,\Delta)\rangle}{F}
\nonumber \nonumber \\
&=& 2 D_\nu \Delta t^{\nu-1},
\end{eqnarray}
which exhibits a dependence on $t$ and $\Delta$ if $\nu\neq 1$ and hence clearly differs 
from $\langle x_0^2(t)\rangle$ that depends only on one time scale.
Ergodicity in the MSD sense is thus broken in this case, 
$\langle\overline{\delta_0^2}\rangle \neq \langle x_0^2\rangle$. 
An example for such a process is the subdiffusive 
continuous time random walk (CTRW) as considered in \cite{HeBur08}.

Conversely, the assumption of ergodicity in the MSD sense clearly implies
\begin{eqnarray}
\langle x^2 \rangle =
\langle \overline{\delta_0^2}\rangle
&=& 2 D_{\nu} \Delta^{\nu}. 
\end{eqnarray}
This in turn violates the GER for time averages,
$\langle \overline{\delta}_F\rangle \neq F\langle \overline{\delta_0^2} \rangle/(2k_B T)$. 
To see this, as in the previous case, we use the GER for ensemble averages Eq. (\ref{einst})
which gives Eq. (\ref{EAdrift0}). 
This in turn results in 
$\langle\overline{\delta_0^2}\rangle=2D_\nu \Delta^\nu\neq 2D_\nu\Delta t^{\nu-1}=2k_B T \langle\overline{\delta_F}\rangle/F$,
unless $\nu=1$.
This case is exemplified by the (Gaussian and ergodic) generalized Langevin systems and fractional
Brownian motion \cite{Deng09}.
For these ergodic cases where the GER for time averages is violated, 
we will later discuss the further generalization of the relation between time averaged drift and MSD.
The above discussion shows that the GER for time averages and/or ergodicity are broken 
for anomalous processes.
When $\nu=1$, both ergodicity and Einstein relation for time averages hold, which constitutes 
a behavior that is unique to normal diffusion.

\paragraph*{Example: Relation of time averaged drift and MSD in the L\'evy walk.} 

We consider a L\'evy walk as follows
\cite{foot}: 
a particle switches the sign of its velocity at random times. 
The sojourn times $0<\tau<\infty$ in a velocity state $+v_0$ or $-v_0$ 
for which the particle does not change its direction
are independent, identically distributed random variables
with a common probability density function (PDF) $\psi(\tau)$.
The initial position at $t=0$ of the particle is $x(0)=0$. We chose to let the particle start
with positive velocity $+v_0$ so that it first travels a distance
$v_0 \tau_1$, after that is displaced $-v_0 \tau_2$, then $+v_0 \tau_3$ and so forth. 
The $\tau_i$ are thereby drawn according to the sojourn time PDF $\psi(\tau)$.
This sojourn time PDF decays like a power-law at large
times, $\psi(\tau) \propto \tau^{-1-\alpha}$. Depending on the specific choice of $\alpha$
this distribution lacks its first moment ($0<\alpha<1$), or the second moment ($1<\alpha<2$). 
In our simulations we will use 
\begin{eqnarray}
\psi(\tau) &=& \left\lbrace \begin{array}{l c c}
\alpha \tau^{-1-\alpha} && \hspace{.5cm} \tau\geq 1 \\
0 && \hspace{.5cm} else\; .
\end{array}\right. \label{psioft}
\end{eqnarray}
Thus in the unbiased case $F=0$ (the biased case will be discussed below), 
the large-time asymptotics reads $\tilde{\psi}(u) \simeq 1 - Au^\alpha$ 
for $0<\alpha<1$ and $\tilde{\psi}(u) \simeq 1 - \langle\tau\rangle u + Au^\alpha$ 
for $1<\alpha<2$
in Laplace domain, with $u\to 0$ the Laplace variable conjugate to $t$. 
Here $A=|\Gamma(1-\alpha)|$, and $\langle\tau\rangle = \alpha/(\alpha-1)$.

The ensemble averaged mean squared displacement for this process is well known \cite{ZumKlaf90}
and yields for $0<\alpha<1$ 
\begin{equation}
\langle x_0^2(t) \rangle = (1 - \alpha) v_0^2  t^2 \, ,\label{msd}
\end{equation}
a quadratic time dependence so that we refer to this regime as the ballistic one.
For $1<\alpha<2$ we have
\begin{eqnarray}
\left\langle x_0^2 (t)\right\rangle &\simeq& 
2 K_\alpha
t^{3-\alpha}, \label{SBmsd}
\end{eqnarray}
whose time dependence is slower than quadratic, but faster than linear so that we call this
regime subballistic (or enhanced diffusion regime). Here we introduced the generalized 
diffusion coefficient, $K_\alpha=v_0^2 A (\alpha-1)/(\langle\tau\rangle\Gamma(4-\alpha))$ \cite{remark}.

For this type of L\'evy walk, depending on $\alpha$ the time averaged MSD Eq. (\ref{tamsd0}) 
can be a random quantity. The properties of the fluctuations of 
$\overline{\delta_0^2}$ were investigated recently \cite{ours, ours2}. 
Its ensemble average yields in particular 
\begin{equation}
\langle\overline{\delta_0^2}\rangle = {1\over |1-\alpha|} \langle x_0^2\rangle , \label{taDel}
\end{equation}
i.e. $\langle\overline{\delta_0^2}\rangle$ differs from the respective
$\langle x_0^2\rangle$ by a factor \cite{ours, ours2, Akimoto12, Ralf13}. 
Note however that in the subballistic case $1<\alpha<2$ the ensemble averaged MSD
depends crucially on the initial conditions: For the equilibrated process that started long before 
the beginning of the measurement we find \cite{Akimoto12, ours, ours2}
\begin{equation}
\langle\overline{\delta_0^2}\rangle = \langle x_0^2\rangle_{eq}, \label{equilib}
\end{equation}
where $\langle\cdots\rangle_{eq}$ denotes the average over an ensemble of stationary processes 
(in contrast to the average $\langle\cdots\rangle$ over an ensemble conditioned on a turning event taking place at $t_0=0$).
Thus, based on the equilibrium MSD one would conclude that the subballistic L\'evy walk is ergodic.
For $0<\alpha<1$ the average sojourn time is infinite. Hence an equilibrated state does not exist
so that strictly speaking weak ergodicity breaking \cite{B92,MarBar05} occurs only in the ballistic case.

Let us further consider a small constant force $F$ acting on the particle of mass $M$.
At each renewal event, the particle starts out at the respective velocity $\pm v_0$
as given in the force-free case discussed above, but then
accelerates according to Newton's law for the duration of the waiting times $\tau$ 
drawn from the PDF Eq. (\ref{psioft}).
This setting leads to a net time averaged drift described by Eq. (\ref{TAdrift}) involving an integral over the 
ensemble averaged drift $\langle x_F\rangle$. The ensemble averaged drift was 
calculated earlier \cite{BarFleu98}. In our particular case we have
\begin{equation}
\langle x_F(t)\rangle
= \left\{
\begin{array}{l l}
\frac{(1-\alpha)F}{2M} t^2 & 0<\alpha<1 \\  \\
\frac{K_{\alpha}F}{Mv_0^2} t^{3-\alpha} & 1<\alpha<2,
\end{array}
\right.
\label{EAdrift}
\end{equation}
which for the subballistic case $1<\alpha<2$ holds also for the drift 
under equilibrium initial conditions, $\langle x_F(t)\rangle=\langle x_F(t)\rangle_{eq}$.
The GER for ensemble averages Eq. (\ref{einst}) holds also for the L\'evy walk,
as is easily verified with Eqs. (\ref{msd}), (\ref{SBmsd}) and (\ref{EAdrift}): 
$\langle x_F \rangle/\langle x_0^2 \rangle = F/(2Mv_0^2)$, where we can assign an 
effective kinetic temperature $k_B T_{eff}/2=Mv_0^2/2$ \cite{remark1}.

\begin{figure}[t]
\centering{
{\includegraphics[width=.4\textwidth]{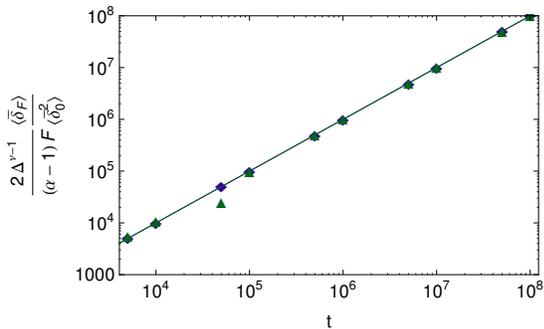}}
}
\caption{\label{NumEina} Ratio 
$(2\Delta^{\nu-1}\langle\overline{\delta}_F\rangle)/((\alpha-1)F\langle\overline{\delta_0^2}\rangle) $ 
for the ballistic case $\nu=2$; $\alpha = 0.5,\;0.7$ 
(blue squares and green triangles, respectively; data points and graphs lie on top of each other).
Symbols represent simulations, solid lines the respective theory Eq. (\ref{genEinst}).
Sample size $5\cdot 10^3$, $v_0=1$, $F=10$, $\Delta=1$.}
\end{figure}

\begin{figure}[t]
\centering{
\vspace{.5cm}
{\includegraphics[width=.43\textwidth]{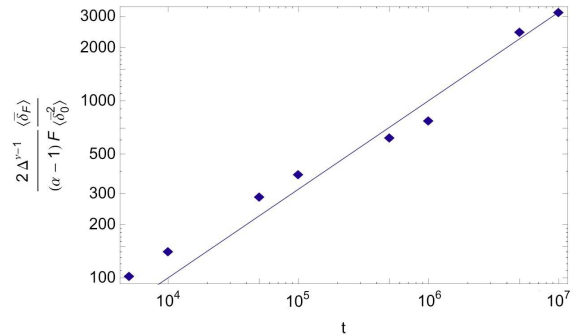}}\\
\vspace{.5cm}
{\includegraphics[width=.43\textwidth]{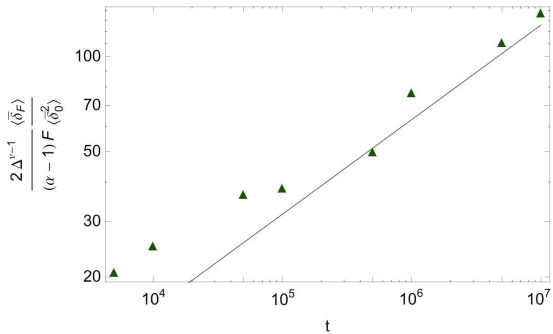}}}
\caption{\label{NumEinb} Ratio 
$(2\Delta^{\nu-1}\langle\overline{\delta}_F\rangle)/((\alpha-1)F\langle\overline{\delta_0^2}\rangle)$ 
for the subballistic case $\nu=3-\alpha$; $\alpha = 1.5$ (upper) and $\alpha = 1.7$ (lower panel). 
Simulational results are depicted as symbols, theory Eq. (\ref{genEinst}) by solid lines.
Sample size $5\cdot 10^3$, $v_0=1$, $F=10$, $\Delta=100$.}
\end{figure}


\begin{table*}[t!]
{\normalsize
\begin{tabular}{l  c c}
 & \parbox[.25\linewidth]{.4\linewidth}{time averaged GER violated \\ \hfill} & \parbox[.25\linewidth]{.4\linewidth}{time averaged GER valid \\ \hfill} \\
\parbox[.25\linewidth]{.15\linewidth}{non-ergodic} &
\fcolorbox{White}{rosa}{\parbox[.5\linewidth]{.4\linewidth}{\hfill\\ ballistic L\'evy walk, \\subballistic L\'evy walk (non-equil.)\\\hfill}}
&\fcolorbox{White}{rosa}{\parbox[.5\linewidth]{.4\linewidth}{ \hfill\\\hfill\\ subdiffusive continuous time random walk \\ \hfill}} \\
\parbox[.25\linewidth]{.15\linewidth}{ergodic} & 
\fcolorbox{White}{rosa}{\parbox[.25\linewidth]{.4\linewidth}{fractional Brownian motion,\\ generalized Langevin equation, \\ subballistic L\'evy walk (equil.)}} 
&\fcolorbox{White}{hellblau}{\parbox[.25\linewidth]{.4\linewidth}{\hfill\\\hfill\\ Brownian motion \\ \hfill}}
\end{tabular}
}
\caption{No-go Theorem: For any anomalous process where the GER for ensemble averages holds, 
ergodicity (in the MSD sense) and GER for time averages never hold both concurrently.
Ergodicity and time averaged GER hold only for normal Brownian motion. 
\label{tb1}}
\end{table*}

In normal Brownian diffusion, the Einstein relation (\ref{einst}) holds also for the time averages,
a trivial observation following from ergodicity, i.e. equality of time and 
ensemble averages for long observation times.
For the L\'evy walk, using Eqs. (\ref{TAdrift}) and (\ref{EAdrift}), 
the ensemble averages of the time averaged drift yield
\begin{equation}
\langle\overline{\delta}_F(t,\Delta)\rangle
= \left\{
\begin{array}{l l}
\frac{(1-\alpha)F t \Delta}{2M} & 0<\alpha<1 \\ \\
\frac{K_{\alpha} F t^{2-\alpha}\Delta}{Mv_0^2} & 1<\alpha<2 \; .
\end{array}
\right.
\label{EATAdrift1}
\end{equation}
These results differ from their corresponding ensemble average $\langle x_F\rangle$.
In the ballistic case $0<\alpha<1$, we have a linear instead of a quadratic dependence
on the time lag. 
In the subballistic case $1<\alpha<2$ the dependence on the lag $\Delta$ is also linear, as compared 
to $\langle x_F\rangle \propto t^{3-\alpha}$. 
In both cases we find a dependence on two time scales, $t$ and  $\Delta$. 
In particular, the time averaged response to an external bias is increasing with $t$.
Hence, in the ballistic regime and in the subballistic regime under nonequilibrium initial conditions
the GER Eq. (\ref{TAeinst}) and ergodicity are violated, compare Eqs. (\ref{EATAdrift1}) and (\ref{taDel}). 
For equilibrium preparation, the subballistic case is ergodic (see Eq. (\ref{equilib})), 
but the time averaged GER Eq. (\ref{TAeinst}) is violated.

Using Eqs. (\ref{tamsd0}), (\ref{taDel}) and (\ref{EATAdrift1}), we can finally write for $0<\alpha<2$
and nonequilibrium initial conditions
\begin{eqnarray}
\frac{\langle \overline{\delta}_F(t,\Delta) \rangle}{\langle \overline{\delta_0^2}(t,\Delta) \rangle} 
&=& \frac{|1-\alpha|F}{2Mv_0^2} \left(\frac{t}{\Delta}\right)^{\nu-1} \label{genEinst}
\end{eqnarray}
with $\nu=2$ in the ballistic and $\nu=3-\alpha$ in the subballistic case. 
This establishes the relationship between dispersion
and drift in the present model. 
The limiting case $\alpha \to 2$ where the sojourn times in the velocity states possess
first and second moment re-establishes the normal diffusion case, so that after sufficiently many collisions or 
changes of direction we have a stationary state. 
Hence in this limit from Eq. (\ref{genEinst}) the well known Einstein relation 
Eq. (\ref{TAeinst}) is recovered for the time averages (where we assign $Mv_0^2/2 = k_B T/2$).
Generally, however, care has to be taken with interpretation of relation Eq. (\ref{genEinst})
in the sense of a fluctuation-dissipation relation, since 
$\langle\overline{\delta_0^2}\rangle$ does not necessarily only comprise thermal contributions.
Conversely, the long excursions leading to ballistic or subballistic anomalous transport may require 
nonthermal energy input that keeps the system out of equilibrium, as is put into effect e.g. 
by motor proteins in subballistic transport within living cells \cite{Caspi02}. This being said, 
the question of the generality of such a time dependent drift-fluctuation ratio Eq. (\ref{genEinst})
arises. We will come back to that later.

Figs. \ref{NumEina}, \ref{NumEinb} show the numerically obtained ratio of the ensemble average drift under the small force $F$, 
$\langle\overline{\delta}_F\rangle$ and the mean of the time averaged MSD, $\langle\overline{\delta_0^2}\rangle$.
The numerics corroborate the theoretical predictions Eq. (\ref{genEinst}) quite well, although in the 
subballistic case the convergence was quite poor, especially for smaller $\Delta$. Thus the numerical results shown 
in Figs. \ref{NumEina}, \ref{NumEinb} also rule out the relation Eq. (\ref{TAeinst}) for the L\'evy walk.

We define an effective mobility $\mu_{eff}$ via 
$\langle\overline{\delta}_F(\Delta,t)\rangle/\Delta = \mu_{eff} F$ and find in the ballistic phase
$\mu_{eff} = (1-\alpha)t/(2M)$, while in the subballistic case $\mu_{eff} = K_{\alpha} t^{2-\alpha}/(Mv_0^2)$. 
Thus the effective mobility increases with the total observation time:
linearly in the first and sublinearly in the latter case \cite{remark3}.

Coming back to a more general setting and to the question of how exactly, 
according to our no-go theorem, the GER for time averages is violated if the process is ergodic, 
we now make a stronger assumption:
We suppose that the anomalous diffusion process is ergodic in the MSD sense and 
that the GER for ensemble averages Eq. (\ref{einst}) holds. 
Thus, Eq. (\ref{EAdrift0}) follows immediately from the anomaly, and ergodicity implies 
$\langle x_0^2(t)\rangle = \langle\overline{\delta^2}\rangle = 2 D_{\nu} t^{\nu}$, 
$0<\nu\leq 2$ and $\nu\neq 1$.
Hence in this case
\begin{equation}
\frac{\langle \overline{\delta}_F(t,\Delta) \rangle}{\langle \overline{\delta_0^2}(t,\Delta) \rangle} 
= \frac{F}{2k_BT} \left(\frac{t}{\Delta}\right)^{\nu-1}, \label{genEinst1}
\end{equation}
which resembles Eq. (\ref{genEinst}) up to a factor $|1-\alpha|$. 
In fact, in the subballistic phase of the L\'evy walk this factor accounts 
for the nonequilibrium initial preparation of the system \cite{ours2, Akimoto12} 
which renders the system nonergodic in the MSD sense.  Eq. (\ref{genEinst1}) thus holds for 
the ergodic L\'evy walks under equilibrium preparation.

\paragraph*{Conclusion.}

We provided a rather general no-go theorem for the validity of 
ergodicity and the GER for time averages: 
At least one of the two properties is violated in systems exhibiting anomalous sub- or superdiffusion,
see Table \ref{tb1}.
This no-go theorem is based on the two assumptions that the process exhibits anomalous diffusion
and that the system obeys the GER for the ensemble averages which leads to 
$\langle x_F\rangle\propto t^\nu$, $\nu \neq 1$.
For anomalous diffusion processes the average response $\langle\overline{\delta}_F\rangle$
to an exerted force exhibits clearly a different time dependence than $\langle x_F\rangle$,
which is due to the non-linear time dependence of $\langle x_F\rangle$.

In particular, we investigated the L\'evy walk model.
While the GER in the L\'evy walk holds for ensemble averages, it is
violated for the time averages.
This is in sharp contrast to the subdiffusive CTRW system studied in \cite{HeBur08}
where ergodicity is violated but a time average GER holds.
The time dependent ratio of drift and dispersion in Eq. (\ref{genEinst}) entails a mobility 
effectively increasing (or a decreasing friction) with the measurement time, 
reflecting the active character of the anomalously 
large excursions in the L\'evy walk.
The L\'evy walk constitutes therefore an example for a system where the relation
between time averaged drift and MSD differs considerably from that of mere ensemble averages.
Moreover we have derived a general relation between time averaged drift and MSD 
for ergodic anomalous diffusion processes, Eq. (\ref{genEinst1}).

Many works are devoted to the question on the ratios between fluctuations and drift 
which are used to define effective temperatures \cite{Cugl11, FieSol02, Barr99}. 
In our case the temperature $1/(k_B T) = 2\langle x_F\rangle/(F\langle x_0^2\rangle)$ 
is well defined, however the ratio 
$ 2\langle \overline{\delta_F}\rangle/(F\langle \overline{\delta_0^2}\rangle) = 1/(k_BT)(t/\Delta)^{\nu-1}$
is time dependent.
Depending on the averaging procedure, either time or ensemble averaging, 
we get different effective temperatures for many kinds of anomalous processes.
Thus when discussing the ratios of fluctuations and drift, the method of averaging has to
be specified carefully.
Theories so far focused on ensemble averages, while experiments seem to focus on 
time averaging (or a mixture of both).

\textit{Acknowledgement: This work was supported by the Israel Science Foundation.}


\begin{thebibliography}{99}
\bibitem{foot} Several types of L\'evy walks are considered in the literature, like the jump model 
where longer jumps are penalized by longer waiting times, or the more physical velocity models
as the one considered here.
\bibitem{remark} Thus for this specific model we have $K_\alpha=D_\nu$ with $\nu=3-\alpha$.
\bibitem{remark1} Note that in Einstein relations the drift under a force is put in relation with the MSD 
of the particle without applied force (compare \cite{BarFleu98} and \cite{vulp12}).
\bibitem{remark3} This behavior is in contrast to the subdiffusive continuous time random walk 
where the effective mobility decreases with the measurement time \cite{HeBur08}.
\end{thebibliography}
\end{document}